\PassOptionsToPackage{unicode}{hyperref}
\PassOptionsToPackage{hyphens}{url}
\documentclass[
  11pt,
]{article}
\usepackage{amsmath,amssymb}
\usepackage{iftex}
\ifPDFTeX
  \usepackage[T1]{fontenc}
  \usepackage[utf8]{inputenc}
  \usepackage{textcomp} % provide euro and other symbols
\else % if luatex or xetex
  \usepackage{unicode-math} % this also loads fontspec
  \defaultfontfeatures{Scale=MatchLowercase}
  \defaultfontfeatures[\rmfamily]{Ligatures=TeX,Scale=1}
\fi
\ifPDFTeX\else
\fi
\IfFileExists{upquote.sty}{\usepackage{upquote}}{}
\IfFileExists{microtype.sty}{% use microtype if available
  \usepackage[]{microtype}
  \UseMicrotypeSet[protrusion]{basicmath} % disable protrusion for tt fonts
}{}
\makeatletter
\@ifundefined{KOMAClassName}{% if non-KOMA class
  \IfFileExists{parskip.sty}{%
    \usepackage{parskip}
  }{% else
    \setlength{\parindent}{0pt}
    \setlength{\parskip}{6pt plus 2pt minus 1pt}}
}{% if KOMA class
  \KOMAoptions{parskip=half}}
\makeatother
\usepackage{xcolor}
\usepackage{mathptmx}
\usepackage[a4paper,margin=22mm]{geometry}
\usepackage{booktabs}
\usepackage{longtable}
\usepackage{array}
\usepackage{graphicx}
\usepackage{caption}
\usepackage{amsmath}
\usepackage{setspace}
\usepackage[none]{hyphenat}
\usepackage{fancyhdr}

\raggedright                      % no justification
\usepackage{titlesec}
\titleformat{\section}{\normalfont\large\bfseries}{\thesection.}{0.6em}{}
\titleformat{\subsection}{\normalfont\normalsize\bfseries}{\thesubsection.}{0.6em}{}
\titlespacing*{\section}{0pt}{14pt}{6pt}
\titlespacing*{\subsection}{0pt}{12pt}{4pt}

\ifLuaTeX
  \usepackage{selnolig}  % disable illegal ligatures
\fi
\IfFileExists{bookmark.sty}{\usepackage{bookmark}}{\usepackage{hyperref}}
\IfFileExists{xurl.sty}{\usepackage{xurl}}{} % add URL line breaks if available
\hypersetup{
  hidelinks,
  pdfcreator={LaTeX via pandoc}}

\author{}
\date{}

\begin{document}

\thispagestyle{empty}
\begin{center}
{\large\bfseries Do Bio-Inspired Algorithms Save Energy in Data Centres?}\\[4pt]
{\large\bfseries Separating Server Packing from Migration Search in Virtual Machine Consolidation}\\[14pt]
\end{center}

\textit{Short title:} Separating packing from search in VM consolidation

\vspace{6pt}

Aun Yichiet\textsuperscript{a,*} \quad Woo Yu Hang\textsuperscript{a}

\textsuperscript{a}\hspace{1.5pt}Faculty of Information and
Communication Technology, Universiti Tunku Abdul Rahman, Jalan
Universiti, Bandar Barat, 31900 Kampar, Perak, Malaysia

\textsuperscript{*}\hspace{1.5pt}Corresponding author. Aun Yichiet,
Faculty of Information and Communication Technology, Universiti Tunku
Abdul Rahman, Jalan Universiti, Bandar Barat, 31900 Kampar, Perak,
Malaysia. Email: aunyc@utar.edu.my

\vspace{10pt}
\begin{center}\textbf{Abstract}\end{center}

Energy-efficient virtual machine (VM) consolidation matters as data
centre electricity demand rises. Bio-inspired metaheuristics, including
ant colony optimisation (ACO), particle swarm optimisation (PSO), and
genetic algorithms, have reported substantial energy savings. However,
these improvements are commonly evaluated against conventional
allocation policies that differ in both initial VM placement and
subsequent migration optimisation, making it difficult to determine the
contribution of metaheuristic search. This study reproduces an existing
consolidation implementation in CloudSim Plus and introduces a matched
control that retains the same initial packing and empty-host
deactivation mechanisms but excludes migration search. ACO, PSO, and a
modified genetic algorithm (MGA) are evaluated across two fleet
configurations and four workload regimes, with ten stochastic
replications per condition. The matched control achieves energy savings
of 0.8--81.4\% relative to the spreading baseline, the amount available
falling sharply as load rises, without migrations or incomplete workload
execution. Across 24 comparisons, metaheuristic search increases energy
consumption in 11 cases, produces identical outcomes in seven, and
reduces consumption in six. Of these, ten increases and five reductions
remain statistically significant after Holm correction. Five of the six
apparent reductions coincide with unfinished tasks and disappear after
normalisation by delivered computational work. Only ACO under light
heterogeneous load retains a 3.22-percentage-point improvement (95\% CI
2.96--3.47). These findings demonstrate that initial packing and host
deactivation account for most of the measured energy savings in the
examined implementations, highlighting the importance of matched control
conditions and workload-completion measures for reliable evaluation of
VM consolidation algorithms.


\hypertarget{introduction}{%
\section{1. Introduction}\label{introduction}}

Global data centre electricity consumption reached 415 TWh in 2024 and
is projected to approach 945 TWh by 2030 {[}1{]}. Efficiency gains at
the facility level can no longer absorb this growth, as the
industry-weighted Power Usage Effectiveness has remained near 1.54 for
six consecutive years {[}2{]}, indicating that the overhead layer is
close to its practical limit. Alternative avenues for optimisation can
therefore be explored within the information technology layer itself,
where computational work is assigned to physical machines.

Scope for such optimisation arises from the way server power scales with
load. A powered but idle host draws between 50\% and 70\% of its peak
demand {[}3{]}, so a fleet operating at low utilisation consumes a
disproportionate share of its rated power. Energy is conserved not by
distributing load evenly across the fleet, but by concentrating it onto
fewer hosts and deactivating the remainder. Virtualisation makes such
concentration possible, and the resulting optimisation problem, virtual
machine (VM) consolidation, has become the principal information
technology intervention available for data centre energy management.

The scale at which these policies will be applied is growing fastest
outside the established markets, and Southeast Asia is where the
question is most immediate. In Johor, Malaysia, 38 data centre projects
hold Electricity Supply Agreements totalling 5.9 GW of maximum demand
against an actual drawn load of 405 MW in December 2024 {[}4{]}, so
contracted demand there exceeds realised demand by more than an order of
magnitude. Most of that capacity has yet to be energised, which means
that the consolidation policies governing its eventual efficiency remain
design decisions rather than installed facts.

VM consolidation is formulated as multi-dimensional bin packing, in
which virtual machines with processor, memory and bandwidth requirements
are assigned to hosts of finite capacity so as to minimise the number of
active hosts. The problem is NP-hard, and its solution space grows
combinatorially with fleet size. Exact methods therefore cannot satisfy
the timescales required for online allocation, while constructive
heuristics such as first-fit and best-fit execute in near-linear time
but leave capacity stranded in fragments too small to allocate.

Bio-inspired metaheuristics occupy the interval between these extremes.
They require no gradient information, operate directly on discrete
assignment spaces, and return a usable solution at any point at which
they are interrupted, which suits an allocator that must respond within
a scheduling interval. Their population-based structure further allows
several candidate configurations to be maintained and compared within a
single iteration, providing a mechanism for escaping the local optima
that greedy packing produces. Three families have dominated this
literature, and each corresponds to a different structural feature of
the problem. Ant colony optimisation constructs assignments
incrementally and retains a pheromone memory across iterations, which
suits a problem in which virtual machines are allocated in sequence and
earlier decisions constrain later ones {[}5{]}. Particle swarm
optimisation represents a complete allocation as a position vector and
converges rapidly, although host indices are categorical and must be
recovered through discretisation {[}6{]}. Genetic algorithms recombine
parent solutions through crossover, an operator capable of preserving
well-packed groups of virtual machines as intact units, and therefore
well matched to grouping problems {[}7{]}.

Bio-inspired algorithms have been reported to reduce energy consumption
by as much as 47\% relative to conventional allocation policies
{[}8,11{]}, with others in the same line reporting gains below 15\%
{[}5,10{]}. Those policies, however, differ from the proposed methods in
two respects simultaneously: the rule governing where virtual machines
are initially placed, and the presence of a subsequent search for
beneficial migrations. Only the second is contributed by the
metaheuristic, and a comparison that varies both cannot determine which
is responsible for the observed saving.

This study develops and applies a controlled experimental framework that
distinguishes the energy effect of initial consolidation from the effect
attributable to metaheuristic search, and evaluates whether reported
improvements persist under matched initial conditions and
workload-delivery requirements. An existing consolidation implementation
of ant colony optimisation, particle swarm optimisation and a modified
genetic algorithm is reproduced in CloudSim Plus and evaluated against a
matched control that applies the same initial packing rule and the same
empty-host deactivation behaviour, but performs no migration search. An
artificial bee colony implementation is carried through the same design
as a structural diagnostic. The contrast between the conventional
baseline and this control measures the effect of packing and
deactivation; the contrast between the control and each metaheuristic
measures the effect of search. Cryptographic fingerprints verify that
all policy arms receive identical virtual machine sets and task
assignments, so that neither contrast is confounded by workload
variation. Energy is reported alongside workload completion throughout,
which allows apparent reductions obtained by leaving tasks unfinished to
be identified rather than credited. The framework is applied across two
fleet configurations and four workload regimes, and three
implementation-level properties of the reproduced system are documented
where they bear on the interpretation of published results.

The remainder of this paper is organised as follows. Section 2 examines
the comparison structure of representative studies. Section 3 specifies
the system model and the decomposition. Section 4 describes the
controlled framework and the measured implementation. Section 5 details
the experimental design. Section 6 reports the results, Section 7
discusses them, and Section 8 concludes.

\hypertarget{related-work}{%
\section{2. Related Work}\label{related-work}}

The theoretical basis for consolidation is the observation of Barroso
and Hölzle that servers are not energy-proportional: an idle but powered
machine draws a large fixed fraction of its peak demand, so energy is
conserved by emptying machines and deactivating them rather than by
balancing load {[}3{]}. Beloglazov and Buyya translated this argument
into the adaptive-threshold heuristics and linear power model that most
simulation-based work in this area inherits, including the
implementation reproduced here {[}12{]}. Cambazard et al.~subsequently
formalised the cost structure as bin packing with linear usage costs,
comprising a fixed cost for each opened bin and a cost linear in its
load, and derived lower bounds for this application {[}13{]}. The
matched control introduced in Section 4 measures how closely an ordinary
best-fit heuristic approaches that bound without any search.

The energy-efficient ant colony system of Liu et al.~is the reference
point for ACO applied to static placement {[}5{]}, and the dynamic
consolidation formulation of Ferdaus et al.~is the corresponding
reference for ACO applied to the migration of running virtual machines
{[}8{]}. Genetic and swarm formulations are comparably developed, and
recent work extends them with reinforcement-learning hybrids {[}14{]},
renewable- and carbon-aware objectives {[}15{]}, and container-level
placement {[}16{]}. Baydoun and Zekri published three treatments of this
problem during 2025, including an ACO-initialised PSO hybrid for energy-
and carbon-efficient consolidation {[}9-11{]}. Across this body of work
the proposed methods differ considerably, but their evaluation structure
does not.

Table 1 classifies the experimental sections of three studies in this
line, together with the implementation reproduced here and the control
introduced here, along the properties that determine what a comparison
identifies.

\begin{table}[htbp]
\centering
\caption*{\textbf{Table 1.} Comparison structure of representative energy-aware consolidation studies, the reproduced implementation, and the present study.}
\small
\begin{tabular}{@{}>{\raggedright\arraybackslash}p{0.15\textwidth}>{\raggedright\arraybackslash}p{0.20\textwidth}>{\raggedright\arraybackslash}p{0.24\textwidth}>{\raggedright\arraybackslash}p{0.12\textwidth}>{\raggedright\arraybackslash}p{0.11\textwidth}@{}}
\toprule
\textbf{Study} & \textbf{Comparator(s)} & \textbf{Comparator shares proposed initial placement} & \textbf{Migration search} & \textbf{Completion reported}\\
\midrule
Liu et al. [5] & FFD, RGGA, ACO, MACO, HACOPSO; CPLEX for bounds & Not stated; comparators retain their own published settings & No; static placement & No\\[4pt]
Ferdaus et al. [8] & FFDL1; MMDVMC & Not stated; all arms begin from a randomly generated load-balanced state & Yes & No\\[4pt]
Baydoun and Zekri [11]\textsuperscript{a} & BFD, FFD, UACS, the authors' prior ACO & Not stated & Yes & No\\[4pt]
Reproduced implementation & CloudSim Plus default allocation policy & No; the default policy spreads, the metaheuristics pack & Yes & No\\[4pt]
This study & Spreading baseline and the matched control & Yes, by construction & Yes; disabled in the control & Yes, per run\\
\bottomrule
\end{tabular}

\vspace{3pt}
{\footnotesize\raggedright \textsuperscript{a}\,Characterised from the openly available version of the two-stage ACO-PSO approach, the immediate predecessor of [9], which carries the same two stated innovations. Entries describe the experimental sections of the cited works as published. ``Not stated'' indicates that the information is not given there, and is not an assertion that the conditions differed.\par}
\end{table}

Each surveyed study compares its proposed method against a policy that
differs from it in both respects at once, so the quantity measured in
each is the joint effect of placement and search. The reported outcomes
are energy and, in some cases, service-level violation rates, and
establishing that the arms delivered equal work additionally requires a
completion count, since an energy reduction obtained by leaving work
unfinished is otherwise indistinguishable from one obtained by
consolidating efficiently. Both properties belong to the experimental
designs rather than to the algorithms, and both are removable at the
cost of one additional arm and one additional metric.

This issue is distinct from the incommensurability documented in the
wider methodological literature. Sörensen's analysis of metaphor-driven
algorithm proliferation {[}17{]}, Kudela's demonstration that widely
used benchmark suites carry artefacts affecting substantial bodies of
published comparison {[}18{]}, and the synthesis by Molina et
al.~{[}19{]} each argue that the incentive structure of the field
rewards proposing algorithms more than comparing them, and Mann's survey
states the domain-specific form of the argument, that published variants
of the problem differ sufficiently that reported numbers are not
commensurable between papers {[}20{]}. Incommensurability limits what
can be concluded by comparing two papers; an unmatched control limits
what can be concluded within one. The remedy is correspondingly
narrower, requiring no standard benchmark or shared workload, only that
one arm differ from the proposed method in a single respect.

\hypertarget{system-model}{%
\section{3. System Model}\label{system-model}}

\hypertarget{placement-migration-and-workload}{%
\subsection{3.1 Placement, migration and
workload}\label{placement-migration-and-workload}}

A simulated data centre in this study comprises a fixed set of physical
hosts and a stream of virtual machines that arrive to run on them. Each
host offers processing elements, memory, storage and network bandwidth.
Each virtual machine requests a quantity of each and carries a share of
the user workload, expressed as cloudlets, which are units of
computation measured in millions of instructions.

Two decisions govern where that work runs. Placement occurs when a
virtual machine arrives: the allocation policy selects a host with
sufficient free capacity to accept it. Migration occurs afterwards: the
policy may move a running virtual machine from its current host to
another, which allows virtual machines spread across many lightly loaded
hosts to be gathered onto fewer hosts so that the remainder can be
powered down. Placement is decided once per virtual machine on arrival,
whereas migration is reconsidered at every scheduling interval for the
lifetime of the simulation. Both decisions are taken by one component,
the allocation policy, and both are where the metaheuristics examined
here act: ant colony optimisation, particle swarm optimisation and the
modified genetic algorithm each supply a rule applied at admission
together with a search that proposes migrations at every subsequent
interval. Section 4 specifies the rules and the searches used.

All four resources are reserved at admission. A host accepts a virtual
machine only while its uncommitted memory, storage, bandwidth and
aggregate processing capacity each meet the request, and
\texttt{VmSchedulerTimeShared} enforces the last of these by testing the
host's remaining unallocated MIPS against the virtual machine's full
nominal demand. What time-sharing permits is that the number of virtual
processing elements resident on a host exceed its physical count, since
each virtual element draws a share of the host's aggregate capacity
rather than owning a core. Capacity in MIPS therefore bounds how densely
a host can be packed, while the count of virtual elements does not.

Formally, let \(P\) be the set of physical hosts and \(V\) the set of
virtual machines, with \(C\) the set of cloudlets representing user
workload. An allocation is a mapping \(A : V \rightarrow P\). The
reservation constraint on memory, storage and bandwidth is

\begin{equation}
\sum_{j\,:\,A(\mathrm{VM}_j)=\mathrm{PM}_i} \mathrm{VM}_j^{\,r} \;\le\; \mathrm{PM}_i^{\,r} \qquad \text{for all } i,\; r \in \{m, s, b\}.
\end{equation}

\begin{figure}[htbp]\centering
\includegraphics[width=\textwidth]{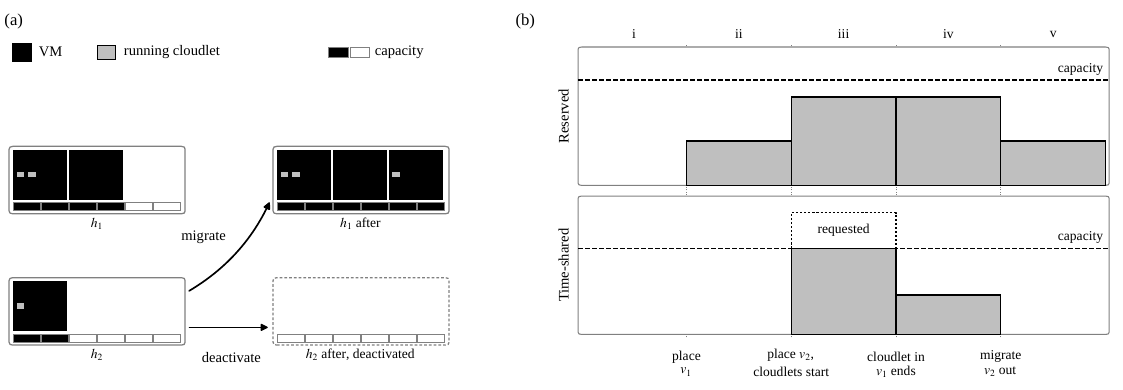}
\caption*{\small\textbf{Fig. 1.} The simulated system. \textbf{(a)} Hosts have finite capacity, drawn as the rail along the foot of each host. A virtual machine is a \emph{reservation} against that capacity, drawn as a block whose width is the share it holds; cloudlets are units of work that execute inside a VM. A running VM may be migrated to another host, and a host left with no VMs is deactivated and stops drawing power. \textbf{(b)} The two resource classes, for one host across five intervals. Memory, storage and bandwidth are reserved when the VM is placed and released only when it leaves: the upper row steps up at each placement, steps down at the migration, and does not move when the cloudlet ends. They cannot be over-committed, so a host whose free memory is below a VM's request is not a candidate for it and the reserved row can never cross the capacity line. Processing is time-shared: the lower row is zero in interval ii, where a VM is resident but idle, rises only while cloudlets execute, and in interval iii shows demand exceeding the host's processing capacity, with the delivered rate capped and the surplus deferred. The fleets are 20 hosts of 256\,GB with 48, 56 or 24 processing elements at 3.0, 2.7 or 2.4\,GHz; the three VM classes reserve 2, 4 and 8\,GB against requests of 1000, 4000 and 9600 MIPS. The contrast between the two rows is what the rest of the paper turns on: an active host draws 300\,W before it does any work and 600\,W at full load, so the dominant term in fleet energy is how many hosts are awake and for how long, which the reservation row sets, rather than how much computing they did, which is the time-shared row. Bar heights in (b) are illustrative proportions chosen for legibility, not measured values from any run.}
\end{figure}

Fig. 1 shows the resulting structure: the host fleet, the virtual
machines resident on each host, the cloudlets they carry, and the two
decision points at which an allocation policy acts.

\hypertarget{power-and-energy}{%
\subsection{3.2 Power and energy}\label{power-and-energy}}

Energy follows from utilisation through the host power model. A powered
host draws a fixed amount regardless of load, plus an amount that rises
with processor utilisation, and it is the size of the fixed term that
makes emptying and deactivating a host worthwhile. The standard linear
form {[}12{]} is

\begin{equation}
P(u) \;=\; k\,P_{\max} + (1-k)\,P_{\max}\,u,
\end{equation}

with \(k\) the idle fraction. This study uses \(k = 0.5\), following the
convention that an idle powered server draws 50 to 70 per cent of peak
{[}3,12{]}. For the Dell PowerEdge R740, with \(P_{\max} = 600\) W, Eq.
(2) gives \(P(u) = 300 + 3.00u\) watts with \(u\) in per cent.

Energy is the time integral of Eq. (2) summed over active hosts,
accumulated at every scheduling tick and reported in kilojoules. The
tick accumulation matters because the alternative, evaluating Eq. (2)
once at a host's mean utilisation and multiplying by its uptime, is
unbiased only if that mean is itself time-weighted. In the simulator
used here a busy host contributes a utilisation sample at most once per
second while an idle host contributes one at every tick, so idle
readings are over-represented and the resulting mean understates true
utilisation by an amount that grows with the idle-but-powered time a
policy produces. Since that quantity is exactly what the arms differ in,
the two estimators disagree by \(-16\%\) to \(+6\%\) depending on arm,
and the tick integral is used throughout. Both quantities are present in
the deposited data {[}21{]}. Absolute energies by arm are given in Fig.
S1.

\hypertarget{decomposition-of-the-energy-saving}{%
\subsection{3.3 Decomposition of the energy
saving}\label{decomposition-of-the-energy-saving}}

The energy saving a policy achieves relative to the baseline is the sum
of two contributions, one from the placement it applies at admission and
one from the search it runs above that placement. Resolving that sum
into its terms requires a third energy measurement, taken under the same
placement rule with the search disabled. Writing \(E_B\) for baseline
energy, \(E_K\) for the control and \(E_X\) for a policy \(X\), all
within the same fleet and load regime,

\begin{equation}
\begin{aligned}
\text{placement effect} &= 100\,(E_B - E_K)/E_B,\\[2pt]
\text{search effect} &= 100\,(E_K - E_X)/E_B.
\end{aligned}
\end{equation}

Both are expressed in percentage points of baseline energy, so they sum
to the policy's total saving. A negative search effect records that the
metaheuristic consumed more energy than the same placement rule with the
search disabled.

\hypertarget{energy-per-unit-of-delivered-work}{%
\subsection{3.4 Energy per unit of delivered
work}\label{energy-per-unit-of-delivered-work}}

Policies complete different numbers of cloudlets, so an energy
comparison alone weighs arms that performed different amounts of work.
Delivered work \(w\) is the fraction of submitted millions of
instructions that were executed, counted as the instruction volume
completed across all cloudlets whether or not the cloudlet itself
reached completion, and energy per unit of delivered work is \(E/w\).
Instruction volume and cloudlet count are reported separately, since a
run can lose a larger share of one than of the other. How strongly to
charge for undelivered work is itself an assumption, so the family

\begin{equation}
E / w^{\gamma}
\end{equation}

is reported at \(\gamma = 0\) (raw energy), \(\gamma = 0.5\), and
\(\gamma = 1\) (undelivered work charged at the policy's own average
rate). For each positive search effect, the flip exponent is the value
of \(\gamma\) at which the gain reverses, and it measures how strong a
charge for undelivered work that gain withstands.

\hypertarget{controlled-experimental-framework}{%
\section{4. Controlled Experimental
Framework}\label{controlled-experimental-framework}}

\hypertarget{separating-admission-from-search}{%
\subsection{4.1 Separating admission from
search}\label{separating-admission-from-search}}

The placement and search effects defined in Section 3 are differences
between three energies measured on one workload. Obtaining them requires
arms that differ from one another by one component at a time, and the
components have to be identified before they can be separated.

An allocation policy in this setting performs two functions. On
admission it chooses a host for each arriving virtual machine.
Thereafter, at each scheduling interval, it decides whether resident
virtual machines move and where they move to. A metaheuristic policy
supplies both: the search it performs operates on an allocation that the
policy's own admission rule produced, and the two arrive together as one
software component. A comparison between such a policy and a framework
default therefore varies the admission rule and the search at the same
time, and the energy difference it yields is the sum of two
contributions in unknown proportion.

The arms used here are organised as three levels of a single stack. The
lowest level admits virtual machines by a rule that spreads them across
the fleet and keeps every host powered. The middle level replaces that
rule with one that packs virtual machines onto as few hosts as capacity
allows and powers down the hosts it empties. The upper level adds a
migration search above the packing. Adjacent levels differ by one
component, so each contrast between them attributes an energy difference
to that component alone.

\hypertarget{policy-arms}{%
\subsection{4.2 Policy arms}\label{policy-arms}}

Five of the six arms occupy the middle and upper levels of the stack and
therefore share one admission rule. Among hosts with sufficient free
memory, storage and bandwidth for the arriving virtual machine, the rule
prefers hosts that are already active, and among those selects the host
with the fewest free processing elements. Hosts whose last virtual
machine departs are powered down, in every arm but artificial bee
colony, whose behaviour in this respect is given in Section 4.5. The
fleet is consolidated at admission, before any search has run, onto as
few active hosts as the capacity constraints permit.

Migration search is the second stage, and it operates on the result of
the first. At each scheduling interval the simulator requests a
migration plan from the allocation policy and applies the
virtual-machine-to-host moves the policy returns. A policy that returns
an empty plan leaves the admission-time placement in force for that
interval.

The control arm, PackOnly, is the shared admission rule with the plan
held empty at every interval. It occupies the middle level of the stack
on its own, and it measures what admission-time packing and host
deactivation contribute together, because the baseline spreads virtual
machines and keeps the fleet powered, so moving to the control
introduces both at once.

The baseline occupies the lowest level. It is CloudSim Plus's default
allocation policy {[}22{]}, which selects, among suitable hosts, the one
with the greatest free capacity. That rule distributes virtual machines
across the fleet rather than concentrating them, and the policy leaves
every host powered for the duration of the run. Table 2 sets out the six
arms by the two properties that distinguish them.

\begin{table}[htbp]
\centering
\caption*{\textbf{Table 2.} The six policy arms, by admission rule, search stage and host-deactivation behaviour.}
\small
\begin{tabular}{@{}llll@{}}
\toprule
\textbf{Arm} & \textbf{Admission rule} & \textbf{Search} & \textbf{Deactivates hosts}\\
\midrule
Baseline & \texttt{VmAllocationPolicySimple} (spreads) & none & no\\
\textbf{PackOnly} & active-first best-fit, shared & \textbf{disabled} & yes\\
ACO & same best-fit rule & ant colony & yes\\
PSO & same best-fit rule & particle swarm & yes\\
MGA & same best-fit rule & modified genetic & yes, once per simulation\\
ABC & same best-fit rule & artificial bee colony & \textbf{no}\\
\bottomrule
\end{tabular}

\vspace{2pt}
{\footnotesize\raggedright\textit{Note.} The baseline is CloudSim Plus's default allocation policy [22], which selects the suitable host with the greatest free capacity and therefore spreads virtual machines across the fleet. Bold marks the properties that distinguish the control and the ABC arm from the other four.\par}
\end{table}

\hypertarget{the-search-stage}{%
\subsection{4.3 The search stage}\label{the-search-stage}}

Ant colony optimisation, particle swarm optimisation and the modified
genetic algorithm are standard implementations of their respective
metaheuristics, differing from one another only in how the migration
plan is constructed.

\begin{figure}[htbp]\centering
\includegraphics[width=\textwidth]{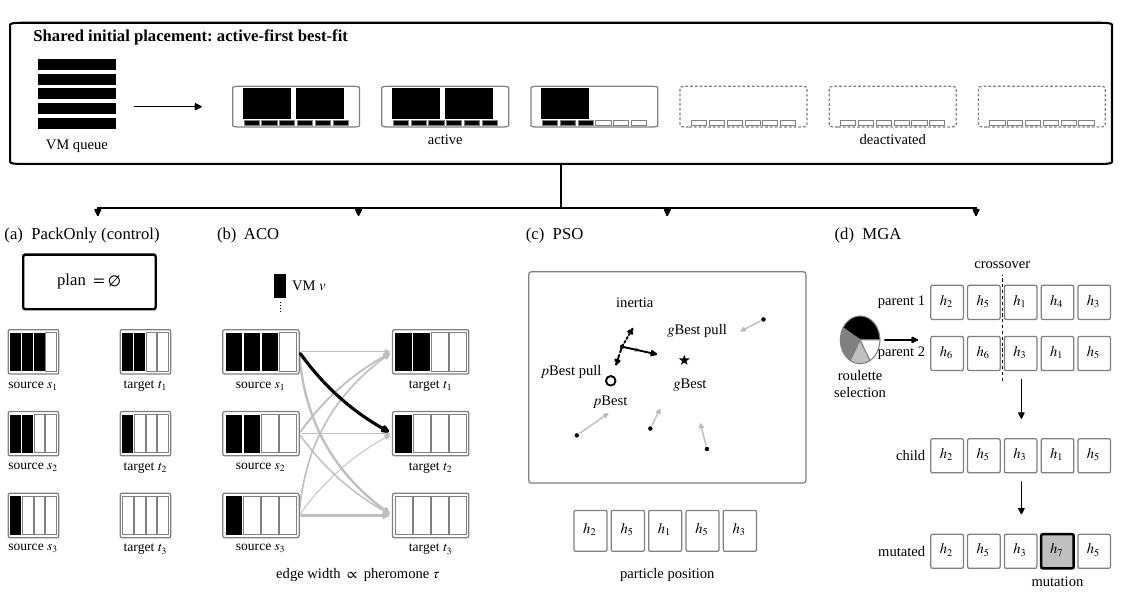}
\caption*{\small\textbf{Fig. 2.} The search stage, and what precedes it. Every arm begins at the band across the top: virtual machines are placed by one rule, which among hosts suitable for the VM prefers those already active and then takes the one with the fewest free processing elements, consolidating the fleet onto as few active hosts as it can and deactivating the rest. This rule is the \texttt{defaultFindHostForVm} override carried, character for character, by all four metaheuristic policy classes, so the arms below branch from an identical starting placement. \textbf{(a)} The control returns an empty migration plan at every scheduling interval, drawn as the same source-to-target question the searches are asked with no edge proposed: the initial placement stands, no VM is migrated and no host is reactivated. \textbf{(b)} Ant colony optimisation treats re-placement as a bipartite graph from source hosts to target hosts, with edge weight proportional to deposited pheromone $\tau$; 5 ants $\times$ 10 iterations $=$ 50 fitness evaluations per invocation, $\tau_0 = 1.0$, $\rho = 0.1$, $\alpha = 1.0$, $\beta = 2.0$, $q_0 = 0.7$, minimising the number of hosts left awake with a tie-break on residual free capacity. \textbf{(c)} Particle swarm optimisation encodes a whole VM-to-host assignment as one particle position, an integer vector over the target-host list, and moves it under inertia and pulls towards the particle's own best and the swarm's best; 10 particles $\times$ 10 iterations $=$ 100 evaluations, $w = 0.5$, $c_1 = c_2 = 2.0$, velocity unclamped. \textbf{(d)} The modified genetic algorithm encodes an assignment as a chromosome, selects parents by fitness-proportionate selection, and applies single-point crossover then mutation; population 10 over 100 generations $=$ 1010 evaluations, mutation rate 0.1, fitness $1/(1 + 0.5 n_{\mathrm{under}} + 0.5 n_{\mathrm{over}})$ against thresholds it derives itself from a control chart over current host utilisations rather than from the SLA contract.}
\end{figure}

Fig. 2 sets out the placement stage that every arm shares, together with
the migration plan each of the three searches computes above it.

Ant colony optimisation represents re-placement as a bipartite graph
from source hosts to target hosts, with edge weights proportional to
deposited pheromone (\(\tau_0 = 1.0\), \(\rho = 0.1\), \(\alpha = 1.0\),
\(\beta = 2.0\), \(q_0 = 0.7\)), and minimises the number of hosts left
awake with a tie-break on residual free capacity. Particle swarm
optimisation encodes a complete virtual-machine-to-host assignment as a
single particle position and moves it under inertia with attraction
towards the particle's own best position and the swarm's best
(\(w = 0.5\), \(c_1 = c_2 = 2.0\), velocity unclamped). The modified
genetic algorithm encodes an assignment as a chromosome, selects parents
by fitness-proportionate selection, and applies single-point crossover
followed by mutation at rate 0.1, with fitness
\(1/(1 + 0.5\,n_{\mathrm{under}} + 0.5\,n_{\mathrm{over}})\) evaluated
against thresholds the algorithm derives from a control chart over
current host utilisations.

Search effort is conventionally reported per invocation, and on that
measure the budgets are 50 evaluations for ant colony optimisation (5
ants over 10 iterations), 100 for particle swarm optimisation (10
particles over 10 iterations), 1010 for the modified genetic algorithm
(an initial population of 10 followed by 100 generations) and 500 for
artificial bee colony. Invocation counts differ by more than three
orders of magnitude between those same policies, which makes effort per
run the comparable unit. Ant colony and particle swarm search at each of
approximately 4130 scheduling intervals in a run, artificial bee colony
searches about nine times under a throttle of one search per 500
simulated seconds, and the modified genetic algorithm searches once.
Section 7 reports effort per run alongside measured search time.

Convergence of a particle swarm depends on the relation between the
inertia weight and the acceleration coefficients, and the Clerc-Kennedy
analysis {[}6{]} gives the parameter region within which particle
trajectories contract towards an attractor. With \(w = 0.5\),
\(c_1 + c_2 = 4.0\) and velocity unclamped, the configuration measured
here sits outside that region, so particle positions remain free to grow
across iterations.

\hypertarget{verification-of-the-design-premises}{%
\subsection{4.4 Verification of the design
premises}\label{verification-of-the-design-premises}}

Two premises carry the design, that the five packing arms share one
admission rule and that every arm receives the same workload, and both
are established by direct inspection of the code.

All four metaheuristic classes override \texttt{defaultFindHostForVm}.
Normalising whitespace and hashing each implementation returns the same
digest, so the rule described in Section 4.2 is carried character for
character by all four, and it is this rule, rather than the framework's
power-aware parent rule, that executes, and the control arm reproduces
the same routine. The metaheuristics therefore inherit a hand-written
best-fit rule rather than the framework's power-aware placement, which
is what the arrangement of the code suggests.

Workload identity is established by a SHA-256 fingerprint computed over
the broker's created-virtual-machine list and the complete
cloudlet-to-virtual-machine assignment. Across 93 fingerprinted runs
spanning all six arms, both fleets, all four load regimes and many
seeds, exactly one fingerprint pair occurs per load regime, and all 272
runs created 100\% of their submitted virtual machines. Every arm
therefore receives an identical virtual machine set in identical
submission order with an identical cloudlet mapping. Submission order is
fixed by construction, so the fingerprint is reproducible across runs
and across machines.

\hypertarget{operating-characteristics-of-the-search-policies}{%
\subsection{4.5 Operating characteristics of the search
policies}\label{operating-characteristics-of-the-search-policies}}

Three characteristics of the policies in this configuration bear on the
results reported in Section 6.

Ant colony, particle swarm and artificial bee colony choose migration
sources by classifying each active host as overloaded or underloaded
against a pair of utilisation thresholds taken from the service-level
contract. The contract in \texttt{CustomerSLA.json} stores those
thresholds as percentages, 20 and 80, and
\texttt{SlaMetricDimension.getValue()} returns them on that scale, while
the host utilisation they are compared against is returned by
\texttt{getCpuPercentUtilization()} as a fraction in \([0, 1]\). Under
that comparison every active host is classified as underloaded, and the
classification returns the same verdict at every interval. All three
policies therefore treat the entire active fleet as a candidate
migration source for the whole run, and migration is driven entirely by
the underload path. The modified genetic algorithm derives its
thresholds from a control chart over current host utilisations and is
unaffected.

The simulator requests a migration plan from each policy at every
scheduling interval, and the modified genetic algorithm answers that
request once. A guard in its allocation routine returns an empty map on
every invocation after the first, so the search runs a single time per
simulation and the placement it produces then stands for the remainder
of the run. In seven of its eight cells the algorithm performs zero
migrations, and its energy in those cells is bit-identical to the
control arm across ten seeds per cell, 70 runs in total.

Consolidation reduces energy only when the hosts a policy empties are
powered down. Ant colony, particle swarm and the modified genetic
algorithm each invoke \texttt{shutdownEmptyHosts} to do so. Artificial
bee colony holds every host powered and reaches its results through
migration alone. Its mean active-host count is 19.98 of 20 in every
cell, matching the baseline to within \(10^{-5}\) of a host.

\hypertarget{experimental-design}{%
\section{5. Experimental Design}\label{experimental-design}}

\hypertarget{configuration}{%
\subsection{5.1 Configuration}\label{configuration}}

The implementation measured here is a research implementation of the
algorithm descriptions cited in Section 2, written for this programme of
work and deposited in full with this paper {[}21{]}. The comparison
therefore targets those descriptions as realised in one codebase, rather
than any particular set of previously reported figures. Table 3
summarises the testbed.

\begin{figure}[htbp]\centering
\includegraphics[width=\textwidth]{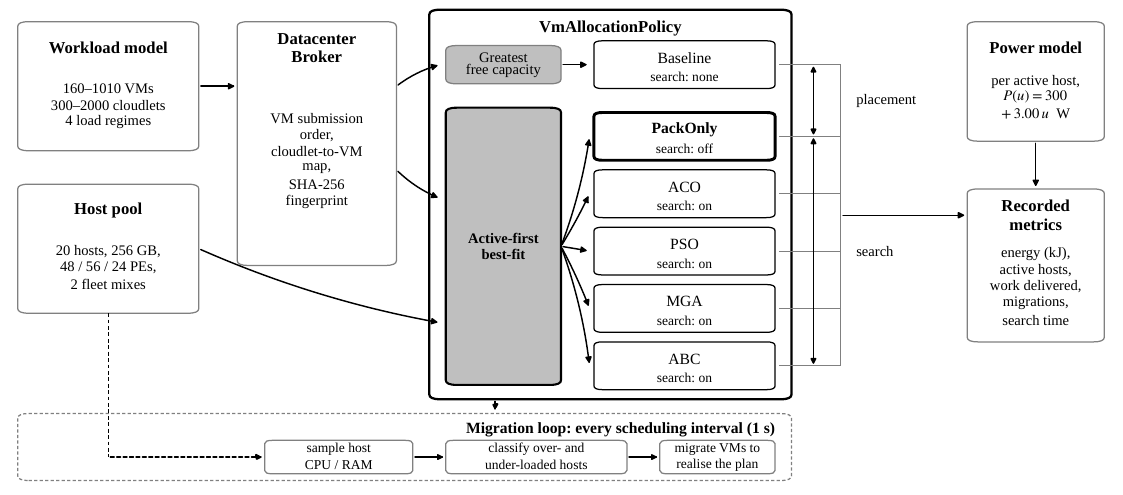}
\caption*{\small\textbf{Fig. 3.} The testbed and the two reported contrasts. The workload model, broker, host pool, power model and migration loop are identical for every arm, and the allocation policy is the only component that changes. The broker submits the same VM set in the same order with the same cloudlet-to-VM mapping to every arm, verified per run by SHA-256 fingerprints of both: across 93 fingerprinted runs spanning all six arms, both fleets, all four load regimes and many seeds there is exactly one fingerprint pair per load regime, and all 272 runs created 100\% of their submitted VMs. The policy slot holds six arms over two placement rules. The spreading baseline uses CloudSim Plus's default placement, which selects the suitable host with the greatest free capacity; the other five share the active-first best-fit rule shown in Fig. 2. PackOnly is that rule with the search switched off, its \texttt{getOptimizedAllocationMap} returning an empty map on every interval. The placement contrast is the baseline against PackOnly and the search contrast is PackOnly against each of the four search policies, both reported in Fig. 4. The fleets are 20 hosts of 256\,GB each, homogeneous as 20 $\times$ 48 processing elements at 3.0\,GHz and heterogeneous as 8 $\times$ 48 at 3.0\,GHz, 6 $\times$ 56 at 2.7\,GHz and 6 $\times$ 24 at 2.4\,GHz; the scheduling interval is 1\,s; the power model is $P(u) = 300 + 3.00\,u$ W per active host, with start-up at 400/450/300\,W for 5\,s and shut-down at 50\,W for 3\,s.}
\end{figure}

Fig. 3 shows the pipeline: the workload model, broker, host pool, power
model and migration loop are common to every arm, and the allocation
policy is the only component that varies across the six.

Two properties of the configuration carry more weight than the rest.
Memory is identical across the two fleets at 5120 GB, so any difference
between them is attributable to the distribution of processing capacity
and power ratings rather than to a change in the constraint that
determines whether a host can accept a virtual machine at all. The
simulated horizon is effectively common to every run. Completion is
governed by stragglers on single-element LIGHT virtual machines, whose
processing allocation is the same under every host fleet and every
allocation policy. Every arm therefore accumulates energy over the same
interval, and the percentage comparisons reported in Section 6 are
percentages of a common quantity.

Replication follows the behaviour of each arm. The baseline and the
control carry no stochastic element and produce bit-identical output in
every recorded column under different seeds at the cells where this was
checked, so each is run once per cell, while the three search policies
are run ten times per cell against recorded seeds. The design therefore
comprises eight fleet-and-load conditions, and repetition within a
condition estimates the variation the search seeds contribute on a fixed
workload rather than sampling further workloads.

\begin{table}[htbp]
\centering
\caption*{\textbf{Table 3.} Fleet composition, workload, replication and instrumentation.}
\small
\begin{tabular}{@{}>{\raggedright\arraybackslash}p{0.27\textwidth}>{\raggedright\arraybackslash}p{0.68\textwidth}@{}}
\toprule
\textbf{Element} & \textbf{Configuration}\\
\midrule
Platform & CloudSim Plus 8.5.5, OpenJDK 21; scheduling interval 1 s\\
Homogeneous fleet & 20 $\times$ Dell PowerEdge R740 (48 cores at 3.0 GHz, 256 GB, 300 W idle / 600 W peak)\\
Heterogeneous fleet & 8 $\times$ R740; 6 $\times$ HPE ProLiant DL380 Gen10 (56 cores at 2.7 GHz, 350/700 W); 6 $\times$ Supermicro 1029U-TR4 (24 cores at 2.4 GHz, 200/400 W)\\
Fleet memory & 256 GB per host, 5120 GB per fleet in both configurations\\
Host transitions & start-up 5 s at 400/450/300 W; shut-down 3 s at 50 W\\
VM classes & three classes on 1/2/4 cores, reserving 2/4/8 GB against 1000/4000/9600 MIPS\\
Load regimes & S1--S4: 160, 380, 760 and 1010 VMs against 300, 750, 1500 and 2000 cloudlets\\
Replication & ten per stochastic cell for ACO, PSO and MGA, each with a recorded seed; baseline and PackOnly deterministic; ABC at one to three replications; 272 runs in total\\
Simulated horizon & makespan spans 0.02\% across all 272 runs (4105.3--4106.2 s)\\
Recorded per run & energy, mean active hosts, processor and memory utilisation, cloudlets submitted and completed, delivered million-instructions, makespan, migrations started, algorithm wall-clock time\\
\bottomrule
\end{tabular}
\end{table}

Recording completion alongside energy separates an energy reduction
obtained by leaving work unfinished from one obtained by consolidating
efficiently, which is the second of the two properties Section 2
identifies as absent from the surveyed designs.

\hypertarget{verification-and-validation}{%
\subsection{5.2 Verification and
validation}\label{verification-and-validation}}

Alongside the two design premises verified in Section 4.4, four
properties of the testbed on which the results of Section 6 depend were
established by direct check rather than assumed.

Determinism in the two reference arms was confirmed directly, repeated
runs of the baseline and the control under different seeds returning
bit-identical output in every recorded column at the cells where this
was checked, which is what allows each to be run once per cell and
treated as a constant in the tests of Section 5.3.

The structural identities reported in Section 6.1 are exact rather than
approximate, since in its seven zero-migration cells the modified
genetic algorithm returns an energy bit-identical to the control across
ten seeds per cell, 70 runs in total, which also confirms that the
control's energy is reproducible from an independent code path.

The energy measure is physically consistent, in that host power is
linear in utilisation and dynamic energy must therefore be proportional
to delivered work at a ratio independent of the policy. Measured across
the homogeneous runs the tick integral of Eq. (2) satisfies this to
within 1.3\%, whereas evaluating Eq. (2) at each host's mean sampled
utilisation gives a ratio varying by two orders of magnitude, which is
the sampling asymmetry set out in Section 3.2.

The capacity floor is reproducible from the configuration, the
per-regime floors of 3, 7, 14 and 16 hosts in the homogeneous fleet and
3, 7, 13 and 18 in the heterogeneous fleet following from the declared
host capacities and the virtual machine mix. Nine of the 272 runs sit
below their floor, and each of those nine delivered less than the full
workload, which is the signature of lost residency rather than of denser
packing.

\hypertarget{statistical-protocol}{%
\subsection{5.3 Statistical protocol}\label{statistical-protocol}}

The control arm is the comparator in every search contrast, and it is
deterministic. Its energy in a given cell is therefore a constant, and
the sampling distribution belongs entirely to the policy arm. Each
contrast is a one-sample \(t\) test of the policy's ten per-seed
energies against that constant, with nine degrees of freedom. The
comparison is paired on the workload as well, since the fingerprinting
described in Section 4.4 establishes that every arm within a load regime
receives the same virtual machines in the same order with the same
cloudlets attached.

Seven of the 24 comparisons are structural identities. The modified
genetic algorithm in its seven zero-migration cells returns an energy
bit-identical to the control across all ten seeds, and those cells are
recorded as \(p = 1\). Restricting the family to the 17 testable
comparisons raises particle swarm optimisation at heterogeneous Scenario
4 to significance, so the family used here reports one benefit fewer
than the narrower alternative would. Table 4 gives the remaining
choices.

\begin{table}[htbp]
\centering
\caption*{\textbf{Table 4.} Statistical procedure.}
\small
\begin{tabular}{@{}>{\raggedright\arraybackslash}p{0.27\textwidth}>{\raggedright\arraybackslash}p{0.68\textwidth}@{}}
\toprule
\textbf{Element} & \textbf{Choice}\\
\midrule
Test & one-sample $t$ against the deterministic control, nine degrees of freedom\\
Family & 24 pre-specified comparisons: \{ACO, PSO, MGA\} $\times$ 2 fleets $\times$ 4 load regimes\\
Correction & Holm--Bonferroni primary; Benjamini--Hochberg $q$-values alongside; simultaneous Bonferroni intervals at 99.79\%\\
Equivalence & two one-sided tests against a $\pm 2$ percentage point margin\\
Cross-arm & Friedman over the eight blocks with Nemenyi post hoc [23], on raw energy and on energy per unit of delivered work, over the five arms excluding ABC and again over the four sharing the packing rule\\
Outside the family & ABC, reported descriptively\\
\bottomrule
\end{tabular}
\end{table}

\hypertarget{results}{%
\section{6. Results}\label{results}}

\hypertarget{decomposition-of-the-energy-saving-1}{%
\subsection{6.1 Decomposition of the energy
saving}\label{decomposition-of-the-energy-saving-1}}

Table 5 separates each policy's saving into a placement component, which
is a property of the cell, and a search component, which is a property
of the policy within that cell. The placement component is the control
arm's saving against the spreading baseline, and it runs from 0.82\% at
heterogeneous Scenario 4 to 81.40\% at homogeneous Scenario 1, obtained
with zero migrations and every submitted cloudlet completed. The
component is therefore strongly load-dependent: under light load the
packing rule captures most of the energy a consolidation policy can
reach, and under the heaviest load tested it captures almost none,
because the fleet at that point admits almost no consolidation at all.

\begin{table}[htbp]
\centering
\caption*{\textbf{Table 5.} Decomposition of the energy saving into placement and search components, in percentage points of baseline energy.}
\small
\begin{tabular}{@{}llrrrr@{}}
\toprule
\textbf{Fleet} & \textbf{Load} & \textbf{Placement} & \textbf{ACO} & \textbf{PSO} & \textbf{MGA}\\
\midrule
Homogeneous & S1 & \textbf{81.40} & $-$0.12$^{***}$ & $-$37.03$^{***}$ & $-$71.62$^{***}$\\
            & S2 & \textbf{58.60} & $-$0.09 & $-$44.32$^{***}$ & 0.00\\
            & S3 & \textbf{24.65} & $+$0.82 & $-$18.04$^{***}$ & 0.00\\
            & S4 & \textbf{11.88} & $-$2.41$^{***}$ & $-$8.06$^{***}$ & 0.00\\
Heterogeneous & S1 & \textbf{67.62} & \textbf{$+$3.22}$^{***}$ & $-$31.08$^{***}$ & 0.00\\
            & S2 & \textbf{43.57} & $+$2.51$^{***}$ & $-$41.70$^{***}$ & 0.00\\
            & S3 & \textbf{13.91} & $-$3.55$^{***}$ & $+$5.90$^{***}$ & 0.00\\
            & S4 & \textbf{0.82} & \textbf{$+$11.76}$^{***}$ & $+$6.61$^{***}$ & 0.00\\
\bottomrule
\end{tabular}

\vspace{2pt}
{\footnotesize\raggedright $^{***}$ significant at Holm-corrected $\alpha = 0.05$ across the 24-comparison family. Bold marks the placement component in every cell, the largest positive search effect, and the one positive search effect that holds after normalisation by delivered work.\par}
\end{table}

Of the 24 search effects, eleven are negative, seven are exactly zero
and six are positive. Fifteen are significant after Holm correction, ten
of them increases in energy and five reductions. Equivalence testing at
a \(\pm 2\) percentage point margin places nine of the 24 within that
band of zero, though seven of those nine are the structural identities
described below rather than statistical equivalences established against
dispersion. Fig. 4 presents the decomposition.

\begin{figure}[htbp]\centering
\includegraphics[width=\textwidth]{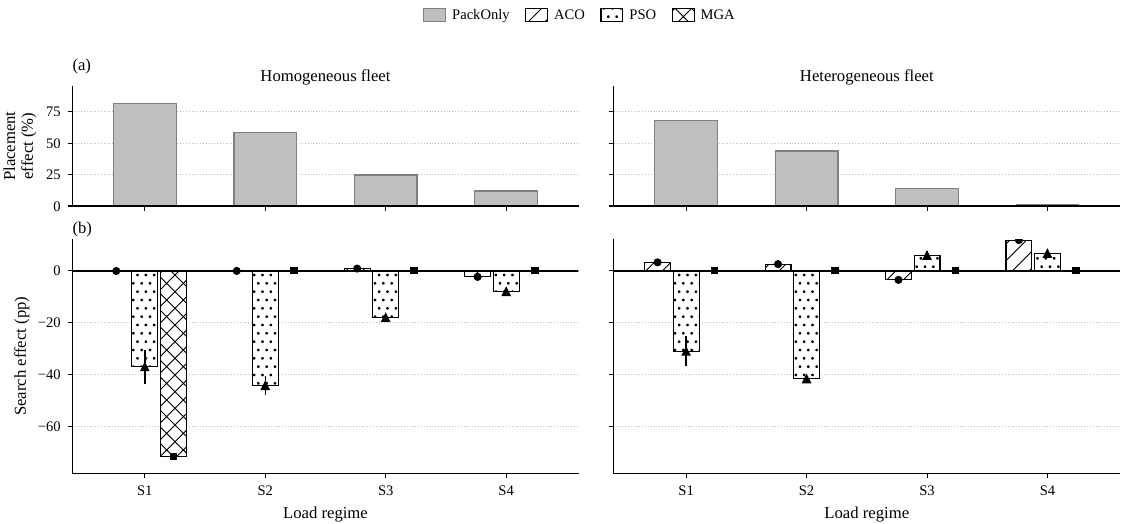}
\caption*{\small \textbf{Fig. 4.} Decomposition of the energy saving into a placement effect and a search effect, by host fleet and load regime. \textbf{(a)} Placement effect: the saving of the PackOnly control against the spreading baseline, that is, the saving obtained by the metaheuristics' own placement rule with the search switched off. It is 81.4, 58.6, 24.6 and 11.9\% at homogeneous S1--S4 and 67.6, 43.6, 13.9 and 0.8\% at heterogeneous S1--S4, and it is obtained with zero migrations and 100\% of cloudlets completed in every cell. \textbf{(b)} Search effect: the saving of each policy minus the placement effect, in percentage points added to it. Bars above the zero rule are cells where the search improved on the placement it was handed; bars below are cells where it reduced part of it. The search effect is negative in 11 of these 24 cells, exactly zero in 7 and positive in 6. The seven zeros are MGA cells in which the search returned no plan at all: MGA runs its search once per simulation and then returns an empty map, so in those cells it reproduces PackOnly. The largest single effect is MGA at homogeneous S1, $-$71.6 pp, the one cell where its search fires, with a mean of 142 migrations, 138 to 146 across the ten runs, taking a 3.01-host placement back up to 17.98 hosts. PSO reduces the placement saving in six of its eight cells ($-$37.0, $-$44.3, $-$18.0, $-$8.1, $-$31.1 and $-$41.7 pp). ACO lies between $-$3.6 and $+$11.8 pp and is not separable from zero after Holm correction in two of its eight cells. Error bars are 95\% $t$ intervals on the mean of ten independent replications. No interval is drawn on the placement bars in (a) or on MGA outside homogeneous S1: PackOnly and the baseline are deterministic, verified bit-identical across seeds and run once per cell, so they have no dispersion to show rather than dispersion too small to see; MGA's ten runs are likewise bit-identical in the seven cells where its search produces no plan. ABC appears in Figs. 6 and S1.}
\end{figure}

The seven exact zeros are the modified genetic algorithm's
zero-migration cells, where its energy is bit-identical to the control
across all ten seeds, so those cells record an identity rather than a
test outcome. In the single cell where that algorithm's search does
fire, homogeneous Scenario 1, it costs 71.62 percentage points, because
a mean of 142 migrations takes a 3.01-host placement back up to 17.98
hosts.

\hypertarget{workload-completion-and-energy-normalisation}{%
\subsection{6.2 Workload completion and energy
normalisation}\label{workload-completion-and-energy-normalisation}}

The control completes every submitted cloudlet in all eight cells, while
the search policies deliver between 81.9\% and 100\% of theirs, which
bears directly on the six positive effects. Table 6 places each against
the work its arm delivered and against the flip exponent \(\gamma\)
defined in Section 3.4, the correction strength at which the gain
reverses.

\begin{table}[htbp]
\centering
\caption*{\textbf{Table 6.} The six positive search effects against delivered work.}
\small
\begin{tabular}{@{}llrrrr@{}}
\toprule
\textbf{Fleet} & \textbf{Load} & \textbf{Policy} & \textbf{Search (pp)} & \textbf{Work delivered} & \textbf{Flip $\gamma$}\\
\midrule
Heterogeneous & S4 & ACO & $+$11.76 & 85.95\% & 0.834\\
Heterogeneous & S4 & PSO & $+$6.61 & 82.83\% & 0.366\\
Heterogeneous & S3 & PSO & $+$5.90 & 81.91\% & 0.356\\
Heterogeneous & S1 & ACO & $+$3.22 & 99.65\% & \textbf{27.7}\\
Heterogeneous & S2 & ACO & $+$2.51 & 91.28\% & 0.496\\
Homogeneous & S3 & ACO & $+$0.82 & 98.49\% & 0.720\\
\bottomrule
\end{tabular}

\vspace{2pt}
{\footnotesize\raggedright\textit{Note.} Bold marks the one effect whose flip exponent lies beyond any plausible charge for undelivered work.\par}
\end{table}

Five of the six reverse at \(\gamma\) between 0.356 and 0.834, so each
is erased by a charge for undelivered work weaker than full, and at full
charging they stand at losses of 0.3 to 12.6 percentage points. The
sixth, ant colony optimisation at heterogeneous Scenario 1, reverses
only at \(\gamma = 27.7\) and therefore holds under any plausible
accounting of unfinished work.

The loss of work tracks migration rather than load, which identifies
where it comes from. All 86 runs that performed no migrations delivered
100\% of their submitted instructions, and every run that fell short of
complete delivery had migrated at least one virtual machine. Across the
sixteen ant colony and particle swarm cells the correlation between
search effect and work lost is \(r = 0.536\) (\(p = 0.032\)), which
points to an arithmetic mechanism rather than an algorithmic one: a host
that stays powered draws its idle term either way, so a policy that
leaves cloudlets unfinished records a lower mean processor utilisation
and therefore a lower energy. Fig. 5 plots every search effect against
the work its arm delivered.

\begin{figure}[htbp]\centering
\includegraphics[width=0.62\textwidth]{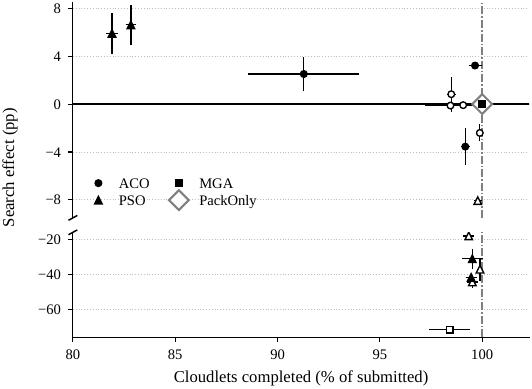}
\caption*{\small \textbf{Fig. 5.} Search effect against delivered work, one point per policy and cell. The control sits at (100\%, 0 pp) by construction, PackOnly being the origin of this plane, and completes 100\% of submitted cloudlets in all eight cells with zero migrations. Five of the six positive search effects sit to its left, between 81.9\% and 98.5\% of work delivered: heterogeneous ACO S4 at $+$11.8 pp with 86.0\% delivered, heterogeneous PSO S4 at $+$6.6 pp with 82.8\%, heterogeneous PSO S3 at $+$5.9 pp with 81.9\%, heterogeneous ACO S2 at $+$2.5 pp with 91.3\% and homogeneous ACO S3 at $+$0.8 pp with 98.5\%. Across the sixteen ACO and PSO cells the correlation between search effect and work lost is $r = 0.536$ ($p = 0.032$). The mechanism is arithmetic rather than algorithmic: the host stays powered either way, so a policy that drops 9--18\% of its cloudlets records lower mean processor utilisation and therefore lower energy. The single exception, and the only clean positive result in the matrix, is ACO at heterogeneous S1: $+$3.22 pp (95\% CI $+$2.96 to $+$3.47) with 0.35\% of work lost. Open markers are the homogeneous fleet, filled markers the heterogeneous fleet; marker shape identifies the policy. MGA's seven zero-search cells coincide exactly with the control and are drawn inside its open diamond. The ordinate is broken: the search effect spans $-$71.6 to $+$11.8 pp and seventeen of the twenty-four cells fall within $\pm$9 pp. No point is clipped. Error bars are 95\% $t$ intervals on both cell means, $n = 10$.}
\end{figure}

\hypertarget{active-hosts-and-the-capacity-floor}{%
\subsection{6.3 Active hosts and the capacity
floor}\label{active-hosts-and-the-capacity-floor}}

The placement effect reported in Section 6.1 follows directly from how
many hosts each rule leaves powered, since an active host draws 300 W
before it performs any work. The baseline holds 19.98 of 20 hosts active
in every cell regardless of load, while the shared best-fit rule packs
virtual machines onto as few hosts as the capacity constraints allow.
Fig. 6 shows the resulting counts against the resource-derived lower
bound.

\begin{figure}[htbp]\centering
\includegraphics[width=\textwidth]{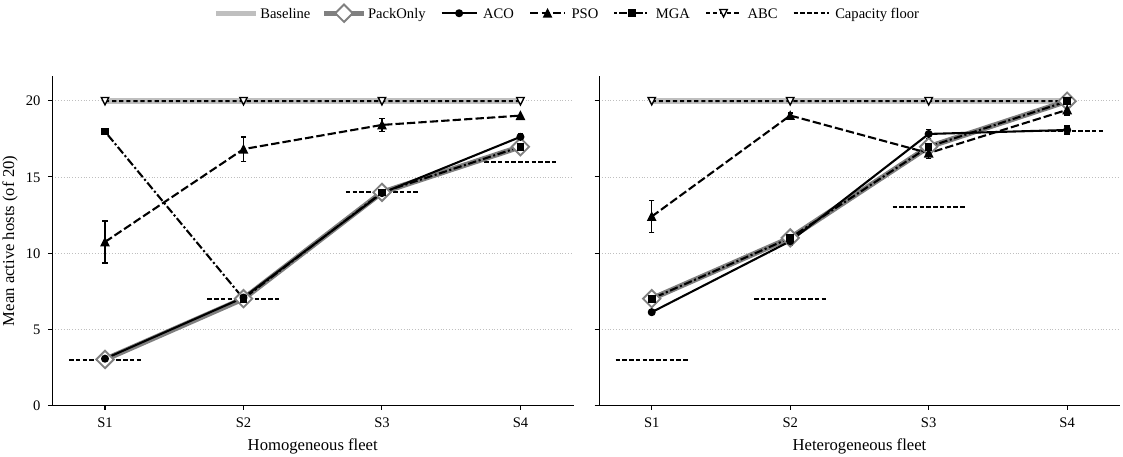}
\caption*{\small \textbf{Fig. 6.} Time-weighted mean active hosts, by arm, host fleet and load regime, against the capacity floor. The floor is the fewest hosts that could hold the regime's requested resources, taking the largest-capacity hosts first: 3, 7, 14 and 16 hosts for the homogeneous fleet and 3, 7, 13 and 18 for the heterogeneous fleet. Processing capacity binds in all eight cells and memory does not, the memory-only floor being 2, 5, 9 and 12 hosts in both fleets since every host carries 256\,GB, so the bound drawn is the processor bound and it is a lower bound on a relaxation rather than an achievable packing. PackOnly reaches it exactly at homogeneous S1, S2 and S3 (3.01, 7.00 and 13.99 hosts against 3, 7 and 14) and sits one host above it at S4. In the heterogeneous fleet it stays two to four hosts above the floor throughout, because the placement rule ranks hosts by free processing elements and is blind to how fast those elements run; this gap is where the one clean positive search effect lies, ACO reaching 6.12 hosts against PackOnly's 7.00 at heterogeneous S1. ABC's line lies on the baseline's in all eight cells, the two agreeing to within $10^{-5}$ of a host, because it holds every host powered. Its energy reduction on the tick integral is correspondingly small, from $-$0.35\% to $+$2.55\%, while continuous migration drops delivered work to 85.6--98.4\% and raises time spent above the 80\% processor threshold above the baseline's in every cell where the baseline registers any. The two reference conditions are drawn as pale ribbons and the four algorithm arms as thin lines over them. Error bars are 95\% $t$ intervals on the mean of ten replications and are drawn only where dispersion exists; PSO at homogeneous S1 ($10.73 \pm 1.38$ hosts) is the widest.}
\end{figure}

That lower bound is set by whichever resource runs out first. Each
virtual machine reserves memory and requests processing elements, so the
fewest hosts that could accommodate a load regime can be computed
separately for each resource, and the larger of the two values is the
one that binds. Memory gives a floor of 2, 5, 9 and 12 hosts across the
four regimes, the same in both fleets because every host carries 256 GB,
while processing gives 3, 7, 14 and 16 hosts in the homogeneous fleet
and 3, 7, 13 and 18 in the heterogeneous fleet. Processing is therefore
the binding constraint in all eight cells.

The control reaches the homogeneous floor exactly at the first three
load levels, placing virtual machines on 3.01, 7.00 and 13.99 hosts
against floors of 3, 7 and 14, so the placement already achieves the
lowest host count available and a search can at best match it. In the
heterogeneous fleet the control settles two to four hosts above the
floor, because the placement rule ranks candidate hosts by the number of
free processing elements and takes no account of how fast those elements
run, filling the slower Supermicro hosts as readily as the faster Dell
ones.

\hypertarget{search-gain-under-heterogeneous-light-load}{%
\subsection{6.4 Search gain under heterogeneous light
load}\label{search-gain-under-heterogeneous-light-load}}

One positive search effect remains significant after correction and
after work normalisation. Ant colony optimisation in the heterogeneous
fleet under light load gains 3.217 percentage points, with a 95\%
confidence interval of 2.962 to 3.472 and a simultaneous Bonferroni-24
interval of 2.735 to 3.699. The test gives \(t(9) = 28.50\) and a
Holm-corrected \(p\) of \(8.2 \times 10^{-9}\). The arm delivers 99.65\%
of its work, and the gain stands at 3.11 percentage points after full
normalisation, with a 95\% confidence interval of 2.82 to 3.41.

The mechanism is the gap identified in Section 6.3 between the control's
placement and the capacity floor. The control sits four hosts above the
floor in that cell, at 7.00 against a floor of 3, and ant colony
optimisation closes part of that gap by reaching 6.12 active hosts. The
condition under which search helps is therefore specific: it helps where
the packing rule has left room, and the homogeneous fleet leaves none.

The size of the result against the rest of the matrix is set by two
further figures. The gain amounts to 4.8\% of the placement effect in
the same cell, and the same algorithm consumes significantly more energy
than the control in two other cells, so one significant positive among
24 pre-specified comparisons supports a claim that search helps under a
particular condition rather than a claim of general advantage.

\hypertarget{friedman-rank-comparison-across-arms}{%
\subsection{6.5 Friedman rank comparison across
arms}\label{friedman-rank-comparison-across-arms}}

Comparing all five arms simultaneously confirms the placement result and
locates where the differences lie. A Friedman test over the eight
blocks, ranking by raw energy, gives \(\chi^2(4) = 18.93\) and
\(p = 8.1 \times 10^{-4}\), against a Nemenyi critical difference of
2.156 rank units. The baseline ranks last at 5.00 and separates from ant
colony optimisation at 2.00, from the control at 2.19 and from the
modified genetic algorithm at 2.56. No other pair separates, and the
three best-ranked arms lie within 0.56 rank units of one another. Under
energy per unit of delivered work the same comparison gives
\(\chi^2(4) = 23.43\) and \(p = 1.0 \times 10^{-4}\), the control ranks
best at 1.56, and it separates from particle swarm optimisation at 4.00
and from the baseline at 4.75.

Removing the baseline puts the paper's own question to the same test.
Across the four arms that share the packing rule, raw energy gives
\(\chi^2(3) = 4.39\) and \(p = 0.223\) against a critical difference of
1.658, so the test does not separate any arm from any other. Under
energy per unit of delivered work the test is significant at
\(\chi^2(3) = 14.89\) and \(p = 0.0019\), the control ranks best at
1.56, followed by the modified genetic algorithm at 1.94 and ant colony
at 2.62, and particle swarm at 3.88 is the one arm from which the
control separates. Fig. 7 shows both rankings.

\begin{figure}[htbp]\centering
\includegraphics[width=\textwidth]{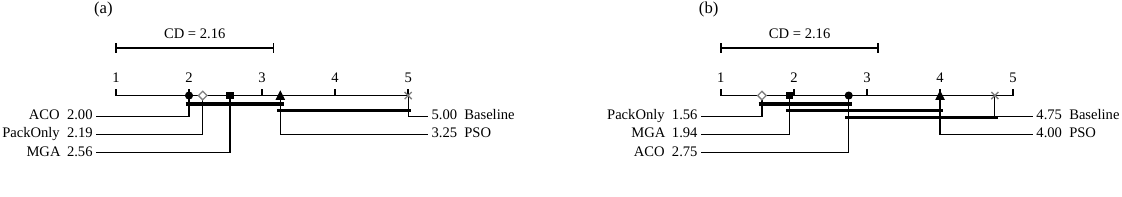}
\caption*{\small \textbf{Fig. 7.} Friedman average ranks with Nemenyi critical difference, over the eight fleet $\times$ load-regime blocks, ranked within block with 1 = lowest energy. \textbf{(a)} Ranked by raw energy: $\chi^2(4) = 18.93$, $p = 8.1 \times 10^{-4}$; CD(.05) = 2.156 rank units. The baseline separates from ACO ($\Delta = 3.00$), PackOnly (2.81) and MGA (2.44), and no other pair separates. \textbf{(b)} Ranked by energy per unit of delivered work: $\chi^2(4) = 23.43$, $p = 1.0 \times 10^{-4}$, and the control at 1.56 separates from PSO at 4.00 and from the baseline at 4.75. Horizontal bars join arms the test cannot separate. With the baseline dropped, which is the paper's own question, the raw-energy test is not significant at all, $\chi^2(3) = 4.39$, $p = 0.223$, CD(.05) = 1.658, while under energy per unit of delivered work the test is significant, $\chi^2(3) = 14.89$, $p = 0.0019$, and PSO is the one arm the control separates from. Eight blocks is below the ten Derrac et al. [23] suggest, and the blocks are two fleets crossed with four monotone load levels rather than independent draws from a population of problems, so the critical difference is correspondingly wide; these tests corroborate the per-cell comparisons of Fig. 4. ABC is excluded; a six-arm sensitivity places it between PSO and the baseline and changes nothing else.}
\end{figure}

\begin{figure}[htbp]\centering
\includegraphics[width=\textwidth]{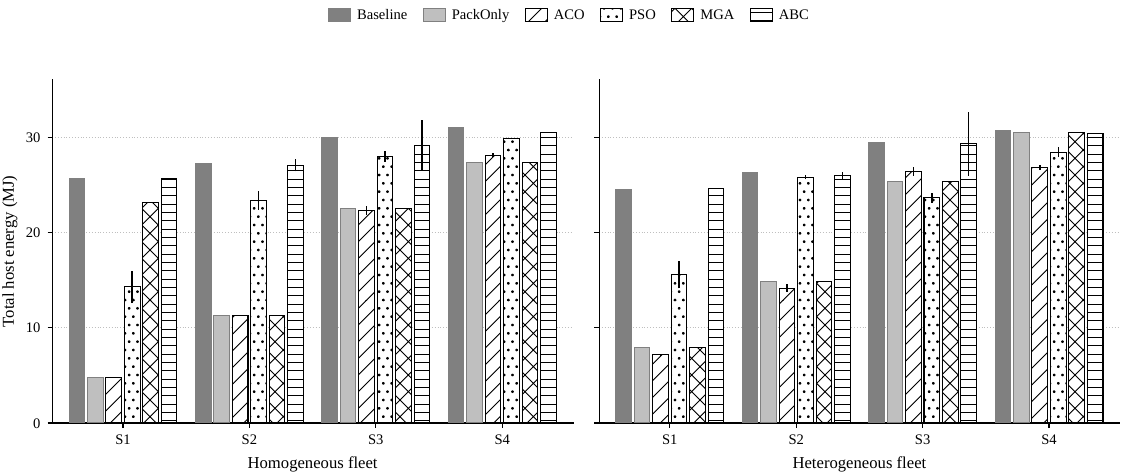}
\caption*{\small \textbf{Fig. S1.} Absolute total host energy by arm, host fleet and load regime, on the tick integral of Eq. (2). The same measurements underlie the percentage decompositions of Figs. 4 and 6, and are given here so that the percentages can be read against the quantities they are taken from. Energy rises with load for every arm and spans 4.77 to 31.02 MJ across the matrix. The spreading baseline sits between 24.49 and 31.02 MJ and varies least with load, because it holds 19.98 of 20 hosts powered in every cell. The control spans 4.77 to 30.43 MJ, the width of that range being the placement effect expressed in joules rather than percentage points. ABC tracks the baseline in all eight cells, within 0.6 MJ, which is the absolute form of the result reported in Section 6.6. Bars are means over ten replications for the stochastic arms and single deterministic runs for the baseline and the control; error bars are 95\% $t$ intervals and are drawn only where dispersion exists. The simulated horizon is common to every run, 4105.3--4106.2 s across all 272 runs, so these totals are accumulated over the same interval and are directly comparable.}
\end{figure}

Once delivered work enters the comparison, the best average rank in both
the five-arm and the four-arm test belongs to an arm that performs no
search, and the arm ranked second performs no migrations. These omnibus
tests run over eight blocks formed by two fleets crossed with four
monotone load levels rather than over independent draws from a
population of problems, and eight blocks is below the ten Derrac et
al.~{[}23{]} recommend, so the critical difference is correspondingly
wide and the tests corroborate the per-cell comparisons of Section 6.1
rather than carrying the argument on their own.

\hypertarget{migration-without-consolidation-and-an-energy-measure-that-concealed-it}{%
\subsection{6.6 Migration without consolidation, and an energy measure
that concealed
it}\label{migration-without-consolidation-and-an-energy-measure-that-concealed-it}}

The artificial bee colony arm migrates continuously and consolidates
nothing, which makes it the clearest case in the matrix of a policy
whose behaviour and whose reported saving come apart. Its mean
active-host count is 19.98 in every cell, identical to the baseline,
because the class holds every host powered. Measured on the tick
integral its saving against the baseline runs from \(-0.35\)\% to
\(+2.55\)\%, which is what a policy that never empties a host should
achieve.

The behaviour shows as migration churn in three measurements that agree.
Delivered work falls to between 85.6\% and 98.4\%, while migration
intensity is highest exactly where the saving is smallest, reaching 7.67
migrations per virtual machine at homogeneous Scenario 1. Time spent
above the 80\% utilisation threshold rises above the baseline's, at
11,125 against 8,315 host-seconds at homogeneous Scenario 4, so a lower
mean utilisation combined with more time above threshold marks work
being moved rather than concentrated. Producing that pattern occupies up
to 1618 s of search per run.

The same arm supplies a direct demonstration of why the energy measure
in Section 3.2 was chosen. Evaluating host power once at each host's
mean sampled utilisation, rather than integrating it over the run,
credits this arm with savings of 1.1\% to 24.1\% against the baseline.
That estimator reports a quarter of the fleet's energy saved by a policy
that holds all twenty hosts powered for the whole run, and it does so
because continuous migration depresses the sampled mean utilisation of
hosts that remain active throughout, which places the saving in the
measurement rather than in the policy. This arm carries one to three
replications per cell, so the observations reported for it are
structural properties of its behaviour rather than distributional
estimates.

\hypertarget{validation-against-corrected-utilisation-thresholds}{%
\subsection{6.7 Validation against corrected utilisation
thresholds}\label{validation-against-corrected-utilisation-thresholds}}

The threshold scale described in Section 4.5 raises a question the
decomposition cannot answer on its own, namely whether the search
effects reported above are properties of migration search or of a
classification step that never fires. A separate arm settles it. The two
policies that consult the service-level thresholds, ant colony and
particle swarm optimisation, were re-run with those thresholds expressed
on the same scale as the utilisation they are compared against, ten
seeds per cell, at the two light-load cells and at the heaviest
heterogeneous cell. The original behaviour is retained as the default
and reproduces the published runs bit-identically, so the two sets of
figures are directly comparable.

Expressing the thresholds on the corrected scale moves the search effect
further from zero rather than closer to it, so the decomposition
reported above is conservative with respect to this design choice. In
the three cells that continue to deliver the full workload, ant colony
falls from \(-0.12\) to \(-49.27\) percentage points at homogeneous
Scenario 1 and from \(+3.22\) to \(-37.76\) at heterogeneous Scenario 1,
while particle swarm falls to \(-81.40\) at homogeneous Scenario 1, a
figure equal to the placement effect in that cell and therefore one that
returns the fleet to baseline energy exactly. The mechanism is visible
in the classification: under the original scale every active host
counted as under-utilised at every interval, so the searches drained
hosts continuously and reached three to six active hosts, whereas a host
on the corrected scale ceases to be a drain candidate once it passes 20
per cent utilisation and consolidation stalls above thirteen. The three
remaining cells deliver between 0.4 and 5.1 per cent of their workload
under the corrected scale, so the comparison there rests on the three
work-conserving cells reported above.

\hypertarget{discussion}{%
\section{7. Discussion}\label{discussion}}

\hypertarget{principal-findings}{%
\subsection{7.1 Principal findings}\label{principal-findings}}

The packing rule accounts for the whole of the measured consolidation
saving. Against the spreading baseline it delivers between 0.82\% and
81.40\% of baseline energy across the eight cells, obtained with zero
migrations and every submitted cloudlet completed. The amount available
is set by load: under light load the rule captures most of what
consolidation can reach, and at the heaviest load tested the fleet
admits almost no consolidation and the rule captures almost nothing.

Enabling migration search above that placement moves energy in both
directions, and more often upward. Across the 24 controlled comparisons
it increases energy in eleven cells, leaves it unchanged in seven and
reduces it in six, with ten increases and five reductions significant
after Holm correction. The search effect lies statistically within
\(\pm 2\) percentage points of zero in nine of the 24, though seven of
those nine are structural identities rather than equivalences
established against dispersion.

Normalising by delivered work removes five of the six reductions. Those
five coincide with between 1.5\% and 18.1\% of the workload left
unfinished, and they reverse at \(\gamma\) between 0.356 and 0.834, each
erased by a charge for undelivered work weaker than full. The one
reduction that holds under every correction applied is ant colony
optimisation under light heterogeneous load, at 3.22 percentage points
with 99.65\% of work delivered.

\hypertarget{the-capacity-floor-and-unfinished-work-as-mechanisms}{%
\subsection{7.2 The capacity floor and unfinished work as
mechanisms}\label{the-capacity-floor-and-unfinished-work-as-mechanisms}}

The findings above raise two questions: why a search operating on an
already-packed fleet so seldom improves on it, and why five of the six
reductions disappear once delivered work is taken into account. The
first is answered by how close the packing rule comes to the fleet's
capacity limit, and the second by how a powered host accumulates energy
whether or not its work completes.

Energy in this model is governed largely by the number of powered hosts,
and the packing rule drives that number close to the lowest value the
resources permit. It reaches the processor-derived floor exactly in
three of the four homogeneous regimes, at 3.01, 7.00 and 13.99 hosts
against floors of 3, 7 and 14. A placement already on the floor offers a
search no better arrangement to find, so the search can only move work
between hosts whose number is already minimal. In the heterogeneous
fleet the same rule settles two to four hosts above the floor, because
it ranks candidate hosts by their number of free processing elements and
takes no account of how fast those elements run. That gap is where the
one positive effect that holds under normalisation occurs, and its size,
4.8\% of the placement effect in that cell, is of the order of the gap
available to it.

Energy accumulates on a powered host whether or not the cloudlets
assigned to it complete, which is the second mechanism. A policy that
leaves work unfinished therefore records a lower mean utilisation and a
lower energy without having improved the allocation, and this accounts
both for the five reductions that reverse under normalisation and for
the positive correlation of \(r = 0.536\) between search effect and work
lost across the ant colony and particle swarm cells.

\hypertarget{implications-for-data-centre-operation}{%
\subsection{7.3 Implications for data centre
operation}\label{implications-for-data-centre-operation}}

The operational content of these results is that a packing heuristic
combined with deactivation of emptied hosts captured, in these
experiments, essentially all of the available consolidation energy,
while requiring no search, completing all submitted work and performing
no migrations. Both components are configuration choices available in
existing allocators rather than research outputs, which places the
saving within reach of an operator specifying a platform rather than
only of one running an optimisation study.

The cost of obtaining those outcomes supports the same reading. Mean
search time per run ranges from 0.4 s for the modified genetic algorithm
on the heterogeneous fleet under light load to 202 s for ant colony
optimisation on the heterogeneous fleet under heavy load, and reaches
1618 s for artificial bee colony, to arrive at outcomes that in
aggregate matched what the control achieved without searching.
Individual runs range more widely still, the longest search occupying
1126 s. Where search did reduce energy without reducing delivered work,
the margin was 3.22 percentage points in one cell of eight.

Section 1 described a build-out in which 5.9 GW of contracted demand
stands against 405 MW drawn, most of it still at the design stage, which
is the window in which a placement-policy result can be acted on. The
consolidation energy available in these experiments comes from a packing
heuristic and a host-deactivation policy, both configuration decisions,
so the lever is available at the point where a platform is specified
rather than only where an optimisation study is run.

Existing guidance addresses the facility layer, with the National
Investment Council recommending a power usage effectiveness at or below
1.4 for hyperscale projects {[}4{]}. A criterion at the information
technology layer, requiring host deactivation and a packing rather than
spreading admission policy, would address a lever that these simulations
show to be strongly load-dependent and, under light load, large.
Measurement on production fleets would establish what it is worth at
deployment scale.

\hypertarget{implications-for-comparative-evaluation}{%
\subsection{7.4 Implications for comparative
evaluation}\label{implications-for-comparative-evaluation}}

A consolidation result obtained by comparing a metaheuristic against a
spreading or first-fit policy, with no matched initial-placement
control, measures the joint effect of the placement change and the
search. It identifies the contribution of the search only where the two
arms share an admission rule, and in the implementations examined here
that contribution was negative more often than positive.

The three operating characteristics reported in Section 4.5 bear on the
same point. A policy whose thresholds are evaluated against values on a
different scale, a policy that searches once and returns an empty plan
thereafter, and a policy that holds every host powered would each be
visible immediately against a matched control, and each records a
plausible saving against a spreading baseline alone.

Four reporting elements follow for comparative placement studies: a
matched initial-placement control arm, delivered work reported alongside
energy, the simulated horizon, and the search budget actually consumed.
Each is available from instrumentation that a simulation study already
produces, and together they cost one additional arm and three additional
recorded quantities.

\hypertarget{extensions}{%
\subsection{7.5 Extensions}\label{extensions}}

The decomposition reported here was carried out under one set of
conditions, and the same design extends past them in three directions.
The fleet in these experiments is twenty hosts, and repeating the
decomposition at rack and multi-rack scale would establish whether the
placement and search components hold their relative sizes once network
topology, thermal coupling and failure domains begin to bear on
allocation. The stochastic arms carry ten replications, set by the
computation available; thirty would narrow the intervals reported here
and bring the non-parametric tests within reach of a plain Bonferroni
threshold. Energy is accounted at the server, and incorporating cooling
and distribution as further parameters would express the saving on a
facility rather than an information technology basis, a difference of
54\% per joule at a power usage effectiveness of 1.54.

\hypertarget{conclusion}{%
\section{8. Conclusion}\label{conclusion}}

This study separated the energy saving obtained by admission-time VM
packing and host deactivation from the additional effect of migration
search. In CloudSim Plus, a matched control retained the reproduced
policies' own placement and host-deactivation behaviour while disabling
search. Across two fleet configurations and four load regimes, with
workload completion recorded in all 272 runs, this design identified
which savings persisted when placement and delivered work were accounted
for.

Packing and host deactivation reduced energy by 0.8--81.4\% relative to
the spreading baseline, without migrations and with every submitted
cloudlet completed, the saving available falling sharply as load rises.
Adding migration search changed the saving by \(-71.6\) to \(+11.8\)
percentage points of baseline energy: energy increased in eleven of the
24 comparisons, remained identical in seven, and decreased in six. Five
of the six reductions coincided with unfinished work and reversed after
normalisation by delivered computational work. Only ACO under light
heterogeneous load retained an advantage: 3.22 percentage points in raw
energy (95\% confidence interval 2.96--3.47), or 3.11 points after full
work normalisation (95\% confidence interval 2.82--3.41), with 99.65\%
of work delivered.

The active-host measurements help explain these outcomes. Packing
reached the processor-derived host-count floor in three homogeneous
regimes, leaving little scope for further consolidation; the surviving
ACO benefit partly closed a gap above that floor in the heterogeneous
fleet. The ABC arm illustrated the complementary risk: it kept the same
number of hosts active as the baseline and left work unfinished, and
under an energy estimator based on mean sampled utilisation it would
have been credited with savings of up to 24.1\%. These findings concern
the implementations and configuration examined, whose threshold handling
and search invocation behaviour materially shaped the results. They
establish the need to isolate search contributions rather than a general
absence of value in metaheuristic consolidation.

The practical finding is that packing and host deactivation captured the
dominant energy saving in these experiments without search or lost work.
The methodological implication is equally direct: comparative
consolidation studies should include the proposed policy's own placement
with search disabled and report delivered work alongside energy,
together with the simulated horizon and actual search effort. These
additions make the incremental contribution of search visible and
prevent reduced workload delivery from being credited as improved energy
efficiency.

\section*{CRediT authorship contribution statement}

\textbf{Aun Yichiet:} Conceptualization, Methodology, Formal analysis,
Writing -- original draft, Writing -- review \& editing, Supervision,
Project administration. \textbf{Woo Yu Hang:} Software, Investigation,
Data curation, Validation, Writing -- review \& editing.

\section*{Declaration of competing interest}

The authors declare that they have no known competing financial
interests or personal relationships that could have appeared to
influence the work reported in this paper.

\section*{Data availability}

The simulation source, configuration, seeds, per-run records, workload
fingerprints and the scripts that regenerate every table and figure are
deposited at Zenodo under doi:10.5281/zenodo.23632346 {[}21{]}, released
under CC BY 4.0.

\section*{Declaration of generative AI and AI-assisted technologies in the manuscript preparation process}

During the preparation of this work, the authors used Claude (Anthropic)
in order to draft and edit the text of the manuscript. The tool was used
under continuous author supervision, sentence by sentence, for language,
structure and exposition. No generative AI tool was used in the design
of the experiments, the configuration or execution of the simulations,
the generation or analysis of the results, or the derivation of the
conclusions reported here; all simulation runs, data and findings are
the authors' own. After using this tool, the authors reviewed and edited
the content as needed and take full responsibility for the content of
the published article.

\section*{References}

{[}1{]} International Energy Agency, Energy and AI, World Energy Outlook
Special Report, IEA, Paris, 2025.

{[}2{]} Uptime Institute, Uptime Institute Global Data Center Survey
2025, Uptime Institute Intelligence Keynote Report 180, July 2025.

{[}3{]} L.A. Barroso, U. Hölzle, The case for energy-proportional
computing, Computer 40 (12) (2007) 33-37. doi:10.1109/MC.2007.443.

{[}4{]} S. Loo, Data Centres, Energy Demand and Sustainability: Can
Malaysia Strike the Right Balance?, ISEAS Perspective 2025/43, ISEAS -
Yusof Ishak Institute, Singapore, 12 June 2025.

{[}5{]} X.-F. Liu, Z.-H. Zhan, J.D. Deng, Y. Li, T. Gu, J. Zhang, An
energy efficient ant colony system for virtual machine placement in
cloud computing, IEEE Trans. Evol. Comput. 22 (1) (2018) 113-128.
doi:10.1109/TEVC.2016.2623803.

{[}6{]} M. Clerc, J. Kennedy, The particle swarm - explosion, stability,
and convergence in a multidimensional complex space, IEEE Trans. Evol.
Comput. 6 (1) (2002) 58-73. doi:10.1109/4235.985692.

{[}7{]} E. Falkenauer, A hybrid grouping genetic algorithm for bin
packing, J. Heuristics 2 (1) (1996) 5-30. doi:10.1007/BF00226291.

{[}8{]} M.H. Ferdaus, M. Murshed, R.N. Calheiros, R. Buyya,
Multi-objective, decentralized dynamic virtual machine consolidation
using ACO metaheuristic in computing clouds, arXiv:1706.06646, 2017.

{[}9{]} A.M. Baydoun, A.S. Zekri, HAPSO: An ACO-initialized,
discretization-aware PSO for energy- and carbon-efficient VM
consolidation in green cloud datacenters, Sustain. Comput. Inform. Syst.
48 (2025) 101258. doi:10.1016/j.suscom.2025.101258.

{[}10{]} A.M. Baydoun, A.S. Zekri, Network-, cost-, and renewable-aware
ant colony optimization for energy-efficient virtual machine placement
in cloud datacenters, Future Internet 17 (6) (2025) 261.
doi:10.3390/fi17060261.

{[}11{]} A.M. Baydoun, A.S. Zekri, Towards efficient VM placement: A
two-stage ACO-PSO approach for green cloud infrastructure, Int. J.
Comput. Netw. Commun. 17 (5) (2025) 117-136.
doi:10.5121/ijcnc.2025.17507.

{[}12{]} A. Beloglazov, R. Buyya, Optimal online deterministic
algorithms and adaptive heuristics for energy and performance efficient
dynamic consolidation of virtual machines in cloud data centers,
Concurr. Comput. Pract. Exper. 24 (13) (2012) 1397-1420.
doi:10.1002/cpe.1867.

{[}13{]} H. Cambazard, D. Mehta, B. O'Sullivan, H. Simonis, Bin packing
with linear usage costs - An application to energy management in data
centres, in: C. Schulte (Ed.), Principles and Practice of Constraint
Programming (CP 2013), Lecture Notes in Computer Science, vol.~8124,
Springer, Berlin, Heidelberg, 2013, pp.~47-62.
doi:10.1007/978-3-642-40627-0\_7.

{[}14{]} A. Ghasemi, G.G. Tejani, A new optimization algorithm based on
the integration of reinforcement learning and HHO algorithm for virtual
machine placement in cloud data centers, Computing 107 (6) (2025) 145.
doi:10.1007/s00607-025-01493-0.

{[}15{]} S.K. Panda, A. Srivastav, A. Singh, K.-C. Li, A renewable
energy-based virtual machine placement algorithm for managing energy and
carbon in geographically distributed datacenters, Cluster Comput. 28
(10) (2025) 666. doi:10.1007/s10586-025-05405-x.

{[}16{]} C. Zou, H. Feng, C. Yuan, S. Pang, GA-CGD: A global
energy-aware container placement strategy for heterogeneous data
centers, Big Data Cogn. Comput. 10 (10) (2026) 330.
doi:10.3390/bdcc10100330.

{[}17{]} K. Sörensen, Metaheuristics - the metaphor exposed, Int. Trans.
Oper. Res. 22 (1) (2015) 3-18. doi:10.1111/itor.12001.

{[}18{]} J. Kudela, A critical problem in benchmarking and analysis of
evolutionary computation methods, Nat. Mach. Intell. 4 (12) (2022)
1238-1245. doi:10.1038/s42256-022-00579-0.

{[}19{]} D. Molina, J. Del Ser, J. Poyatos, F. Herrera, The paradox of
success in evolutionary and bioinspired optimization: Revisiting
critical issues, key studies, and methodological pathways, Swarm Evol.
Comput. 98 (2025) 102063. doi:10.1016/j.swevo.2025.102063.

{[}20{]} Z.Á. Mann, Allocation of virtual machines in cloud data centers
- A survey of problem models and optimization algorithms, ACM Comput.
Surv. 48 (1) (2015) 1-34. doi:10.1145/2797211.

{[}21{]} {[}dataset{]} Y. Aun, Y.H. Woo, Separating server packing from
migration search in bio-inspired VM consolidation: simulation data,
source and analysis, Zenodo, 2026. doi:10.5281/zenodo.23632346.

{[}22{]} M.C. Silva Filho, R.L. Oliveira, C.C. Monteiro, P.R.M. Inácio,
M.M. Freire, CloudSim Plus: A cloud computing simulation framework
pursuing software engineering principles for improved modularity,
extensibility and correctness, in: Proceedings of the 2017 IFIP/IEEE
Symposium on Integrated Network and Service Management (IM), IEEE, 2017,
pp.~400-406. doi:10.23919/INM.2017.7987304.

{[}23{]} J. Derrac, S. García, D. Molina, F. Herrera, A practical
tutorial on the use of nonparametric statistical tests as a methodology
for comparing evolutionary and swarm intelligence algorithms, Swarm
Evol. Comput. 1 (1) (2011) 3-18. doi:10.1016/j.swevo.2011.02.002.

\end{document}